\documentclass{mn2e}
\input{psfig.sty}
\usepackage{mnras_cite}
\newcommand{\apj}{Astrophys. J.}
\newcommand{\aj}{Astron. J.}

\newcommand{\mnras}{Mon. Not. R. Astron. Soc.}

\newcommand{\gal}{{\rm gal}}

\newcommand{\lin}{{}}
\def\sun{\hbox{$\odot$}}

\newdimen\hssize
\newdimen\hdsize 
\begin{document}

\title[z=4 Lyman Break Galaxy Properties]{Where are the $z=4$ Lyman Break Galaxies? 
Results from Conditional Luminosity Function Models of Luminosity-dependent Correlation Functions\thanks{Based on data collected
at Subaru Telescope, which is operated by the National Astronomical Observatory of Japan.}} 
\author[Cooray et al.]{Asantha Cooray$^1$, Masami Ouchi$^2$\thanks{Hubble Fellow}\\
$^1$Center for Cosmology, Department of Physics and Astronomy, University of California, Irvine, CA 92697, USA\\
$^2$Space Telescope Science Institute, 3700 San Martin Drive, Baltimore, MD 21218, USA}

\maketitle
%------------------------------------------------------------------------------

\begin{abstract}
Using the conditional luminosity function (CLF) --- the luminosity distribution of
galaxies in a dark matter halo --- as a way to model galaxy statistics, we
study how  $z=4$ Lyman Break Galaxies (LBGs) are distributed in dark matter halos.
For this purpose, we measure luminosity-dependent clustering of LBGs in the Subaru/XMM-Newton Deep Field by
separating a sample of 16,920 galaxies to three magnitude bins in $i'$-band between 24.5 and 27.5.
Our models fits to data show a possible trend for more luminous galaxies to appear
as satellites in more massive halos; The minimum halo mass in which satellites appear is
$3.9^{+4.1}_{-3.5} \times 10^{12}$ 
$M_{\sun}$,  $6.2^{+3.8}_{-4.9}\times 10^{12}$ $M_{\sun}$, and $9.6^{+7.0}_{-4.6} \times 10^{12}$ $M_{\sun}$ (1-$\sigma$ errors)
for galaxies with $26.5 < i' < 27.5$, $25.5 < i' < 26.5$,  and
$24.5 < i' <25.5$ magnitudes, respectively. The satellite fraction of galaxies at $z=4$ in these magnitude bins
is 0.13 to 0.3, 0.09 to 0.22, and 0.03 to 0.14, respectively, where the 1$\sigma$ ranges account for differences coming
from two different estimates of the $z=4$ LF from the literature.
To jointly explain the LF and the large-scale linear bias factor of $z=4$ LBGs as a function of rest-UV luminosity requires
central galaxies to be brighter in UV at $z =4$ than present-day galaxies in same dark matter mass halos. Moreover,
UV luminosity of central galaxies in halos with total mass greater than roughly $10^{12}$ M$_{\sun}$  
must decrease from $z=4$ to today by an amount more than the luminosity change for galaxies
in halos below this mass. This mass-dependent luminosity evolution is preferred at more than
3$\sigma$ confidence level compared to a pure-luminosity evolution scenario where all galaxies decrease
in luminosity by the same amount from $z=4$ to today. The scenario preferred by the data is consistent
with the ``down-sizing'' picture of galaxy evolution.
\end{abstract}


\section{Introduction}
The conditional luminosity function  (CLF; Yang et al. 2003b, 2005), or the
luminosity distribution of galaxies within a dark matter halo of mass $M$, $\Phi(L|M,z)$,  
captures an important property that determines how the large scale structure
galaxy distribution is related to that of the dark matter at a given redshift $z$. The CLF approach has been used
to show why the galaxy luminosity function (LF) has the shape captured by the
Schechter (1976) form with $\Phi(L) \propto (L/L_\star)^\alpha \exp(-L/L_\star)$
(Cooray \& Milosavljevi\'c 2005b; see, also, Benson et al. 2003) and to derive
certain statistical properties of the galaxy distribution as a function of the environment 
and redshift (Cooray 2005b).

The CLFs extend the analytical halo model  (see, Cooray \& Sheth 2002 for a review)
by dividing the mean number of galaxies as a function of the halo mass, or the so-called halo occupation number $N_g(M)$ 
to a distribution in galaxy luminosity such that $\Phi(L|M)=dN_g(M)/dL$
and using $\Phi(L|M)$, the CLF, to model observed statistics rather than $N_g(M)$ itself.
In our models, the CLF is further divided to two parts involving central galaxies and satellites in dark matter halos.
The central galaxy CLF is taken to be a log-normal, while the satellites are described with a power-law
distribution in luminosity (Cooray \& Milosavljevi\'c 2005b; see, also, Yang et al. 2003b; Zheng et al. 2005).
When one studies statistics such as the group or cluster luminosity function,
one is directly measuring the CLF as appropriate for that mass scale (Cooray \& Cen 2005).
Using the CLF approach, one studies not only the clustering statistics, which are at the
two-point or higher level in the galaxy distribution, but also one-point statistics such as distribution functions.
While the halo occupation number has been the preferred option to describe galaxy clustering
(e.g., Benson et al. 2000; Peacock \& Smith 2000; Seljak 2000; Scoccimarro et al. 2001; Berlind \& Weinberg 2002; Berlind et al. 2003; 
Kravtsov et al. 2004; Zehavi et al. 2005; Zheng et al. 2005),  applications with CLFs are numerous. 

For example, in Cooray (2005c), CLF models of galaxy clustering was used to extract certain properties of
the galaxy distribution with clustering statistics measured in Sloan Digital Sky Survey (SDSS; York et al. 2000),
COMBO-17 (Phleps et al. 2005), DEEP2 (Coil et al. 2004), GOODS (Lee et al. 2005), and from Subaru/XMM-Newton Deep Field (Ouchi et al. 2005). Similar studies, primarily based on the 2dF (Colless et al. 2001) data are described in Yang et al. (2003, 2005).
Recent numerical work on the galaxy distribution (Conroy et al. 2005), 
primarily based on the assumption that each galaxy is associated with a subhalo, supports
analytical models of the galaxy distribution. On the other hand, the suggested
age dependence in halo bias (e.g., Gao et al. 2005; Wechsler et al. 2005; Zhu et al. 2006; Harker et al. 2006) 
is expected to affect analytical models since current models of galaxy statistics are taken to be only a function of halo mass.
The additional dependence related to the halo age, and reflected in terms of an environmental dependence in data, is
expected to affect parameter estimates at a level, at most, around 10\% (Zheng \& Weinberg 2005; Cooray 2006).

While high signal-to-noise ratio galaxy clustering statistics as a function of the galaxy luminosity exists at low
redshifts from surveys such as SDSS (Zehavi et al. 2002; Zehavi et al. 2005) and 2DF (Norberg et al. 2001; Norberg et al. 2002), 
clustering measurements subdivided to galaxy properties are limited at redshifts greater than unity due to small sample sizes of
LBGs recovered in drop-out imaging  surveys
(e.g., Steidel et al. 1998; Adelberger et al. 2003; Adelberger et al. 2005; Giavalisco \& Dickinson 2001). In the context of analytical models,
these clustering measurements mostly around $z \sim 3$  have been used to study the halo occupation number describing
the number of LBGs within dark matter halos at corresponding redshifts (Mo \& Fukugita 1996; Baugh et al. 1998; Governato et al. 1998;
Mo, Mao \& White 1998; Bullock et al. 2002; Moustakas \& Somerville 2002; Hamana et al. 2004). 

Moreover,  galaxy clustering bias factors for LBGs have been  used to set a
halo mass scale for these galaxies (e.g., Giavalisco \& Dickinson 2001; Wechsler et al. 2001;
Porciani \& Giavalisco 2002; Adelberger et al. 2005). 
These estimates seem to have suggested a higher mass for LBG hosting halos than indicated by dynamical measurements
(Pettini et al. 2001). This discrepancy has now led to concepts such as merger-bias (e.g., Scannapieco \& Thacker 2003; 
Furlanetto \& Kamionkowski 2005). It could also be that previous LBG mass measurements based on clustering bias were
affected by assumptions related to the shape of the clustering correlation function (such as a power-law shape), 
statistical limitations related to low LBG sample sizes, or both.

While  previous measurements were limited by statistics of LBGs, wide-field
imaging surveys from instrument such as Suprime-Cam on Subaru (e.g., Miyazaki et al. 2002)
are now slowly increasing the number of drop-out galaxies (Ouchi et al. 2001) with
clustering and LF measurements still continuing today as increasing samples of high redshift galaxies are slowly recovered.
In the case of Subaru Deep Fields, the number of LBGs exceeds twenty thousand in a contiguous square degree field, which is
about 10 to 100 times larger than previous surveys both in the number of sources and the survey area.
The luminosity dependence of $z=4$ LBG 
clustering was first shown in Ouchi et al. (2004b; see, also, Allen et al. 2005; Hildebrandt et al. 2005)
using the first set of imaging data in the Subaru Deep Field (SDF; Kashikawa et al. 2004).
Recently, Kashikawa et al. (2006)  extended the luminosity-dependent clustering measurements using $\sim$ 4500 LBGs
in the same field by adding additional deeper imaging data (down to $i \sim 27.5$)  to the original
GTO images used in Ouchi et al. (2004a, 2004b). All these measurements now show a clear departure from a power-law correlation function at non-linear scales
below ten arcseconds (Ouchi et al. 2005; also Lee et al. 2005). Such a departure is
expected from analytical models based on the halo occupation  number (Zehavi et al. 2004; Zheng 2004).
The sample size of LBGs in Ouchi et al. (2005) at $z=4$ is adequate enough to consider multiple bins in luminosity
to make independent measurements of the luminosity-dependent clustering, instead of the luminosity averaged single
correlation function measured in initial surveys.
Such luminosity-dependent clustering measurements, when combined with CLF models and the LBG LF,
could improve mass estimates for LBG hosting dark matter halos and address the implied discrepancy between
halo mass implied by clustering bias and the dynamical estimate  (Scannapieco \& Thacker 2003).

\begin{figure*}[t]
\centerline{\psfig{file=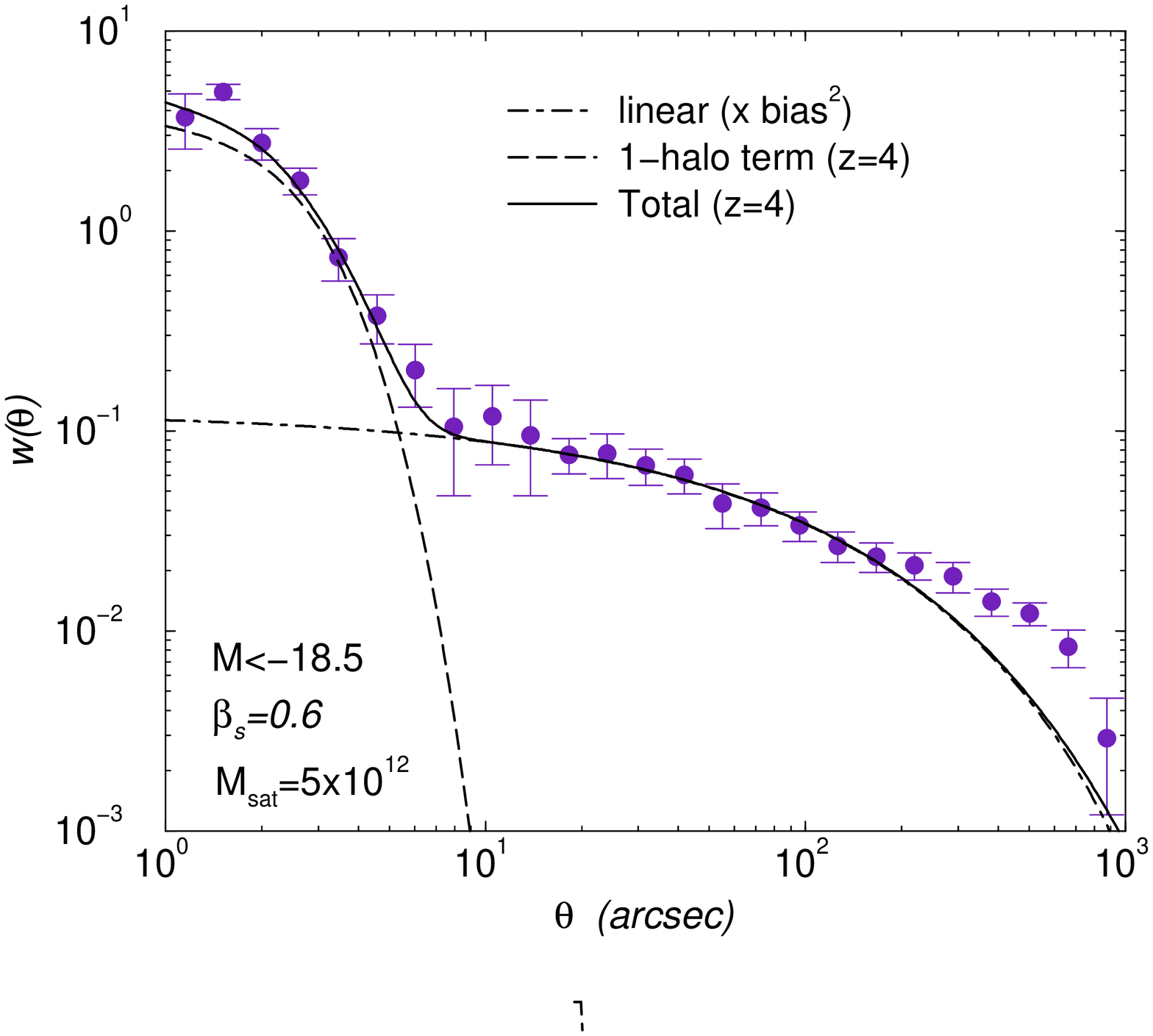,width=\hssize,angle=-90}
\psfig{file=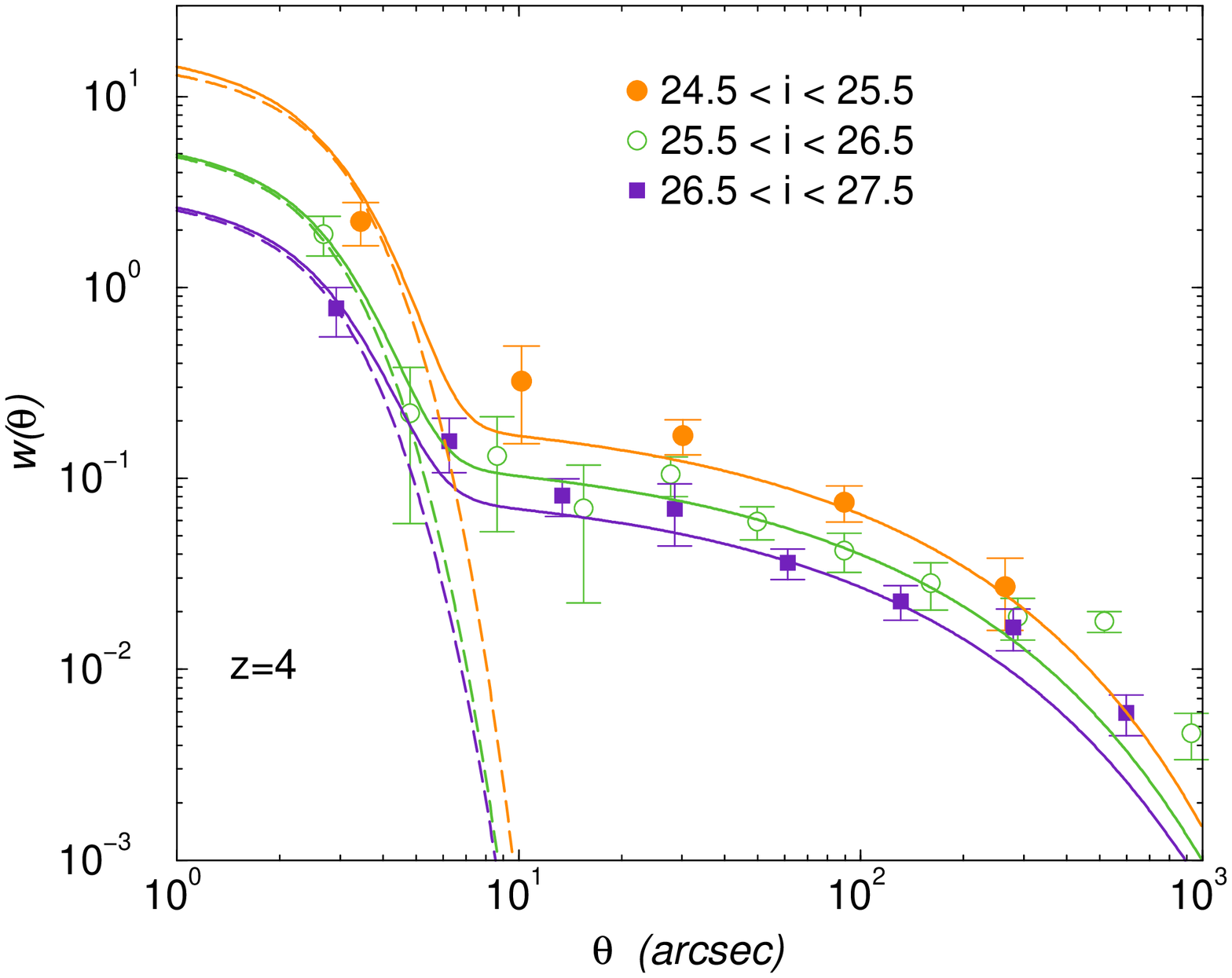,width=\hssize,angle=-90}}
\caption{Angular correlation function of 
LBGs at $z \sim 4$ as measured in the Subaru/XMM-Newton Deep Field. {\it Left panel:}  Clustering of galaxies with $i$-band magnitudes brighter than 27.5,
corresponding to rest-frame $M_B <-18.5$, as previously published in Ouchi et al. (2005). 
The dot-dashed line shows the prediction based on linear theory at $z \sim 4$, scaled by the large-scale
bias factor for galaxies with $M_B <-18.5$, while the dashed line is the
1-halo term with parameters $\beta_s$ and $M_s$ for the satellite distribution (see, equation~3) 
The errors here only include sample variance; Cosmic variance due to finite field size is expected to increase errors at angular scales
greater than 300$''$ and the implied disagreement between measurements and the model will be
reduced. {\it Right panel:} Galaxy clustering at $z \sim 4$ from the Subaru/XMM-Newton Deep Field
as a function of the observed i-band magnitude. Due to neglect of cosmic variance, when model fitting the data, we ignore measurements
at angular scales greater than 300$''$.}
\end{figure*}

In this paper, we will use the same LBG sample as the one discussed in Ouchi et al. (2005).
This sample comes from the Subaru/XMM-Newton Deep Field 
and involves $\sim$ 17,000 LBGs at $z \sim 4$ based on color criteria over an area of
a square degree.
The redshift distribution for a subset of this LBG sample has been estimated based on a 
Monte Carlo analysis combined with spectroscopic observations (Ouchi et al. 2004a). 
In Ouchi et al. (2005) the luminosity dependence of $z=4$ LBG clustering 
was established as a function of the increasing faint-end magnitude of the galaxy sample while the bright-end was kept the same. 
Here we make a new set of measurements where clustering is determined in 3 luminosity bins that are independent of each other
as such measurements are easy to model fit with a likelihood analysis using CLF models. In the case
of original overlapping luminosity bins in Ouchi et al. (2005), the covariance between
bins with a threshold magnitude must be included in a proper model fitting analysis of the data, especially
when comparing estimated parameters as a function of luminosity. Establishing the required
overlapping covariance, however, is challenging in practice.

We measure the clustering of galaxies within three luminosity bins, in observed $i'$-band magnitudes between 24.5 and 27.5 in one magnitude wide bins, 
using the same procedure as described in Ouchi et al. (2005). 
Using autocorrelation functions at angular scales
greater than 10$''$, we derive the large-scale linear bias factor for $z=4$ LBGs, as a function of the LBG luminosity.
Since there is a large variation in the faint-end
estimates of the $z=4$ LBG LF, we make use of two LFs published by Sawicki \& Thompson (2005) and Ouchi et al. (2004a). The
Ouchi et al. (2004a) LF suggests a steep slope at the faint-end, while the Sawicki \& Thompson LF suggests a flat slope.
The CLF model fits to clustering measurements and LFs are then used to extract general properties of the LBG population
at $z \sim 4$.  These include the fraction  of LBGs that appear as satellites in
dark matter halos and the mass scale at which LBGs begin to appear as satellites at a given luminosity. 
 
The paper is organized as follows: In the next  section, we briefly outline luminosity-dependent
clustering measurements in the Subaru/XMM-Newton Deep Field. A detailed description of
the data and the analysis procedure is described in Ouchi et al. (2005), where clustering was
measured as a function of the faint-end of the observed magnitude; Here, we have updated the analysis to
consider independent bins in magnitude or luminosity and to avoid overlapping bins in  luminosity. Note, however, that uncertainties
in the redshift distribution for the LBG sample may lead to small overlaps between luminosity bins; We
 we ignore such complications in the analysis and the interpretation.
In Section~3, we  outline basic ingredients in the empirical model related to CLFs
and how galaxy clustering statistics can be derived from them. We refer the reader to Cooray (2005c)
for a more detail description of this approach and an initial comparison of
models to observed clustering measurements.  In Section~4, we extract parameters related to 
$z \sim 4$ LBG CLFs and provide a detailed discussion of our results based on model fits.
We conclude with a summary of our main results and implications related to the connection between LBGs and dark matter halos
in Section~5. Throughout the paper we assume a cosmological model with currently favored parameters of
$\Omega_m=0.3$, $\Omega_\Lambda=0.7$, a scaled Hubble constant of  $h_{\rm 70}=1$  in units of 70 km s$^{-1}$ Mpc$^{-1}$ 
unless otherwise noted explicitly, and a normalization of the
matter power spectrum today at 8 $h^{-1}$ Mpc scales of $\sigma_8=0.85$, unless otherwise stated explicitly.

\section{LBG Clustering as a function of luminosity at $z \sim 4$}

To measure luminosity dependent clustering of LBGs, we make use of imaging data 
from the Subaru/XMM-Newton Deep Field. The data and the analysis procedure related to autocorrelation function measurements
are described in Ouchi et al. (2005) and we provide a brief summary here (full details related
to the clustering measurement will be published elsewhere; Ouchi et al. in preparation, Hamana et al. in preparation).
The $z \sim 4$ LBG sample involves 16,920 galaxies over an area of one square degree.
The sky distribution  of these galaxies is shown in Figure~1 of Ouchi et al. (2005). The
redshift distribution of these LBGs is taken to be same as the  one used in Ouchi et al. (2005).
This distribution was originally established in
Ouchi et al. (2004a) with a combination of spectroscopic measurements and
Monte Carlo simulations. These LBGs peak at a redshift of 4 but the distribution extends $\pm$ 0.5 in redshift and
drops-off rapidly thereafter at the two ends. The contamination of this sample  has been estimated to be 5\%.

The galaxy clustering in this sample of LBGs was measured using the Landy \& Szalay (1993) estimator 
with random samples of 100,000 sources. The bootstrap errors are estimated using the method of Ling et al. (1986). 
The  5\% contamination in the redshift distribution is expected
to change the angular correlation function of galaxies by about 10\%. In model fitting the data, however,
we ignore such a small correction to the correlation function.  While in Ouchi et al. (2005)
clustering is presented as a function of the faint-end magnitude, while the bright-end is kept the same,
here we divided the sample to four bins: From 23.5 to 27.5 in $i'$-band magnitudes with bins of one magnitude wide.
We measured the autocorrelation function of galaxies in each of the four luminosity bins, but  due to
low sample statistics  in the brightest luminosity bin between $23.5 <i' < 24.5$
the autocorrelation function is only measured with an overall signal-to-noise ratio less than unity. 
Thus, we do not model fit LBG clustering in the brightest bin here.

\begin{figure}
\centerline{\psfig{file=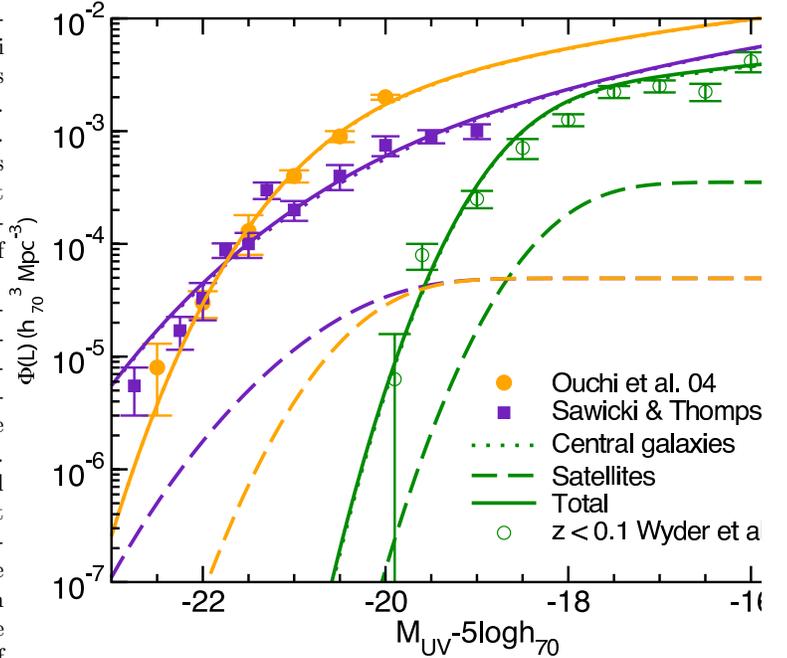,width=10cm,angle=0}}
\caption{
The LF of LBGs at $z\sim 4$. We show two descriptions: from Ouchi et al. (2004a) involving a steep
slope at the faint-end and Sawicki \& Thompson (2005) involving a flat slope at the faint-end.
For comparison, we also show the $z<0.1$ Far-UV
LF from GALEX (from Wyder et al. 2005), whose wavelength corresponds well with the rest-frame band of the $z \sim 4$ 
LBG sample. The curves show best-fit model descriptions of the LF based on CLF models.
The dotted, dashed, and solid
lines show the contribution to the LF from central galaxies, satellites, and the total, respectively.}
\end{figure}

\begin{figure}
\centerline{\psfig{file=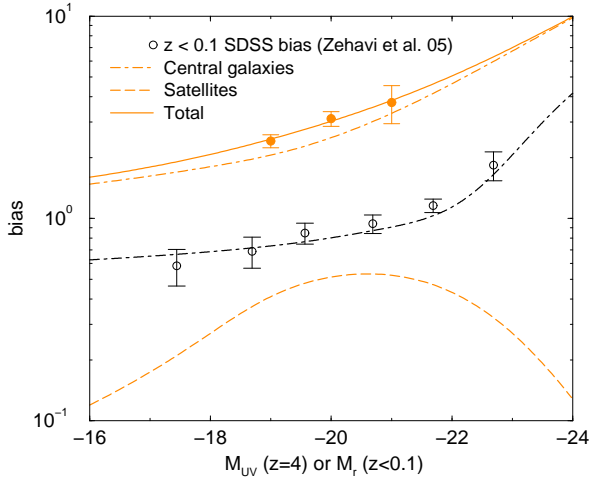,width=7.8cm,angle=-90}}
\caption{
The large-scale galaxy bias as a function of the galaxy luminosity. The $z\sim4$
bias factors are derived using the autocorrelation function at angular scales greater than 10$''$
as a function of the luminosity. For reference, we also show $z<0.1$ bias factors
as a function of the galaxy  luminosity at rest $r'$-band luminosity from SDSS (Zehavi et al. 2005).
Note the general increase in the $z\sim4$ bias factor relative to values found today. 
Due to complications associated with comparing two samples at two different rest wavelengths, we make
no model comparison on the redshift evolution of the bias factor here.}
\end{figure}

Fig.~1 summarizes the autocorrelation function measurements. In Fig.~1({\it left panel}), we show clustering of all galaxies down to
$i < 27.5$, while in the {\it right panel} we show clustering in three luminosity bins in $i'$-band magnitudes
between 24.5 and 27.5, at steps of unity.
Due to uncertainties associated with estimating errors, we ignore measurements with angular separations greater than 300$''$
when model fitting the data. At such large separations, due to the finite size of the survey, 
cosmic variance effects start to become important and the estimated errors here ignore increase in errors related to the finite
field of view. To understand the extent to which these large angular scale measurements 
could affect our conclusions we also included them in one of our model-fit runs.
With clustering measurements as a function of the luminosity bin our best-fit results
did not change significantly though we found out that parameter constraints increase by 15\% to 20\% due 
to the corresponding increase in $\chi^2$ values with the addition of few extra data points at large angular separations.
Such an improvement, however, is artificial as we ignored cosmic variance errors here.

\section{Galaxy Clustering Based on Conditional Luminosity Functions}

In order to construct the luminosity-dependent LBG correlation function at $z \sim 4$, we follow 
Cooray \& Milosavljevi\'c (2005b) and Cooray (2005c) 
to define the redshift-dependent CLF, $\Phi(L|M,z)$,
giving the average number of galaxies with luminosities between $L$ and $L+dL$ that 
resides in halos of mass $M$ at a redshift of $z$. As in previous applications, the CLF is separated 
into terms associated with central and satellite galaxies,  such that
\begin{eqnarray}
\Phi(L|M,z)&=&\Phi_{\rm cen}(L|M,z)+\Phi_{\rm sat}(L|M,z) \nonumber \\
\Phi_{\rm cen}(L|M,z)  &=& \frac{1}{\sqrt{2 \pi} \ln(10)\sigma_{\rm cen} L} \times \nonumber \\
&& \quad \quad \exp \left\{-\frac{\log_{10} [L /L_{\rm c}(M,z)]^2}{2 \sigma_{\rm cen}}\right\}  \nonumber \\
\Phi_{\rm sat}(L|M,z) &=& A(M,z) L^{\gamma(M)}\,.
\label{eqn:clf}
\end{eqnarray}

In Equation~1, $L_c(M,z)$ is the relation between central galaxy luminosity  and its parent halo mass,
 taken to be a function of the redshift, while $\sigma_{\rm cen}$, with a fiducial value of 0.17 (as
measured with SDSS in Cooray 2005c), is the  log-normal
dispersion in this relation.  
For an analytical description of the  $L_{\rm c}(M,z)$ relation, we follow Cooray \& Milosavljevic (2005a)
and write
\begin{equation}
\label{eqn:lcm}
L_{\rm cen}(M,z) = L_0(1+z)^{\alpha} \frac{(M/M_1)^{a}}{[b+(M/M_1)^{cd(1+z)^{\beta}}]^{1/d}}\, .
\end{equation}

In Fig.~2, we summarize two different estimates on the $z=4$ LBG LF from the literature (from Ouchi et al. 2004a and Sawicki \& Thompson 2005), 
with the two estimates showing large differences on the faint-end slope  while the bright-end density remains the same.
At $z \sim 4$, the rest-wavelength of Subaru observations corresponds to a wavelength of 1500 $\AA$ and
to establish the $L_{\rm cen}(M,z)$ relation for rest UV-band, we make use of the $z<0.1$ LF from Wyder et al. (2005)
using GALEX data in the Far-UV band.  Our conclusions do not differ significantly even if we used the Near-UV band
LF of Wyder et al. (2005). This $z< 0.1$ LF is also shown in Fig.~2. Following the approach of Vale \& Ostriker (2004), the parameters 
that best describe this LF are $L_0=5.7\times10^{9} L_{\sun}$, $M_1=10^{11} M_{\sun}$,
$a=4.0$, $b=0.57$, $c=3.85$, and $d=0.23$ (with $\alpha$ and $\beta$ both zero). 
We assume the overall shape given by this relation holds for all LBGs, though it must be modified with appropriate values of
$\alpha$ and $\beta$ to describe the high-redshift  LF. This approach was used in Cooray (2005b)
to describe the redshift evolution of the rest B-band LFs from low redshifts to a redshift of 3.
Here, we consider $\alpha$ and $\beta$ as free parameters to be determined from the data and
use the $z=4$ LBG LFs, from Ouchi et al. (2004a) and Sawicki \& Thompson (2005),
as well as the LBG bias factor as a function of the luminosity
to establish these two parameters. 

For satellites, the normalization $A(M)$ of the satellite CLF can be obtained by defining 
$L_{\rm sat}(M,z)\equiv L_{\rm tot}(M,z)-L_{\rm cen}(M,z)$ and requiring 
that $L_{\rm sat}(M,z)=\int_{L_{\rm min}}^{L_{\rm max}} \Phi_{\rm sat}(L|M,z)LdL$ where
the minimum luminosity  of a satellite is $L_{\rm min}$.
In the luminosity ranges of interest, and due to the numerical value chosen below for the slope $\gamma$, 
our CLFs are mostly independent of the exact value assumed for $L_{\rm min}$ as long as it below the
minimum luminosity probed by observations.  To describe the total luminosity of a halo, 
we make use of the following phenomenological form:
\begin{eqnarray}
L_{\rm tot}(M,z) = \left\{\begin{array}{ll}
L_{\rm cen}(M,z)  & M \leq M_{\rm sat}\\
L_{\rm cen}(M,z)\left(\frac{M}{M_{\rm sat}}\right)^{\beta_s(z)} & M>M_{\rm sat}
\end{array}\right. 
\label{eqn:ltot}
\end{eqnarray}
Here, $M_{\rm sat}$ denotes the mass scale at which satellites begin to appear in dark matter halos
with luminosities of those in the given bin and $\beta_s(z)$ 
is the power-law slope in the luminosity, in addition to the slope of the
total luminosity--halo mass relation.
We will constrain these parameters from clustering data, as a function of the LBG luminosity, 
and use those constraints to determine the mass scale in which LBGs appear as satellites and
other parameters such as $\beta_s$.

While the above form refers to the total luminosity, when $L_{\rm tot}(M,z) > L_{\rm cen}(M,z)$, 
this total luminosity
 must be distributed over a number of satellite galaxies in the halo when describing the satellite CLF. 
We take a power-law
luminosity distribution and set  $\gamma(M,z)=-1$ in Equation~1 based on previous results derived on the CLF of
galaxy groups and clusters (Cooray \& Milosavljevi\'c 2005b) and direct
measurement in clusters such as Coma where $\gamma=-1.01^{+0.04}_{-0.05}$. It is
not clear if the same slope exists at high redshifts.
At low redshifts, $\gamma$ is a function of halo mass and changes abruptly at the low-luminosity end
when one is studying the dwarf galaxy population (Cooray \& Cen 2005), though the latter is not likely to be an issue
as the galaxy sample here is brighter than the luminosity scale of dwarfs. Thus, while the choice of $\gamma\sim -1$
is motivated by the cluster LF at low redshifts, we considered other values and found out that
 setting $\gamma$ to a value  between -0.7 and -1.3, over the set of parameter values we tested, did not change our results 
significantly.  Other parameters related to
the satellite CLFs are described in Cooray (2005c) and the results derived  here
 are mostly insensitive to numerical values for these parameters since variations here only lead to
small changes to the overall CLF. The two main parameters  of our model are $M_{\rm sat}$ and $\beta_s$.
We explicitly determine them from the data here and show constraints imposed by clustering measurements.

In  our description, the central galaxy CLF takes a log-normal form while the satellite galaxy CLF takes a power-law form in luminosity.
Such a separation describes the LF best with an overall good fit to the data in the K-band as explored by
Cooray \& Milosavljevi\'c (2005b) and 2dFGRS $b_J$-band in Cooray (2005).  Our motivation for a log-normal distribution also comes
from measured galaxy cluster LFs that include bright central galaxies where 
a log-normal component, in addition to the Schechter (1976) form, is required to fit the data (e.g., Trentham \& Tully 2002). 
Similarly, the stellar mass function as a function of halos mass in semi-analytical models
is best described with a log-normal component for  central galaxies (Zheng et al. 2005).
The scatter we suggest for the $L_c(M,z)$ relation is also reflected in other statistics of the galaxy
distribution in groups and clusters such as the ``luminosity-gap'' statistic (e.g., Milosavljevi\'c et al. 2005).
If central galaxy luminosities are assigned to dark matter halos with the scatter ignored, biases
could be easily introduced. As discussed in Cooray (2005b),
a one-to-one assumption between mass and luminosity for LBGs, could account for
 the  suggested difference between $z=3$ LBG halo masses based on clustering bias interpretations and
spectroscopic measurements.

The overall shape of the LF is {\it strongly} sensitive to the shape of the $L_{\rm c}$--$M$ relation including
the scatter, and less on details related to the $L_{\rm tot}$--$M$ relation or the satellite CLF.
The non-linear part of the galaxy correlation function, or any clustering statistic, probes the satellite distribution
and constraints can be put on the $L_{\rm tot}$--$M$ relation.  Thus, the combination of one-point statistics
such as the LF and two-point statistics such galaxy clustering could separately aid in measuring the overall CLFs
of LBGs at $z \sim 4$.

In previous discussions of galaxy clustering under the halo model
the occupation number has been widely used as a way to relate statistics of dark matter to galaxies 
(e.g., Kauffmann et al. 1999; Benson et al. 2000; Berlind et al. 2003;
see, review in Cooray \& Sheth 2002). To compare with models of the halo occupation number, CLFs can be easily 
integrated such that
\begin{eqnarray}
N_{\rm cen}(M,z) &=& \int dL \, \Phi_{\rm cen}(L|M,z) \nonumber \\
N_{\rm sat}(M,z) &=& \int dL \, \Phi_{\rm sat}(L|M,z) \, .\nonumber \\
\end{eqnarray}
Since the halo occupation number captures no information on the luminosity distribution of galaxies,
models involving the halo occupation number cannot be used to model the galaxy LF easily.  
In the case of satellites,  $N_{\rm sat}(M,z) = A(M,z) \int dL\, L^{\gamma(M)} g_{\rm sat}(L|M,z)$ where
\begin{equation}
g_{\rm sat}(L|M,z) = \frac{1}{2}\left[1+{\rm erf}\left(\frac{\log(L_{\rm cen}(M,z)/2.0)-\log(L)}{\sigma_{s}}\right)\right] \, ,
\end{equation}
with $\sigma_{s}=0.3$; Such a model description forces satellites  to have a luminosity distribution 
with the brightest satellite below the luminosity of the central galaxy.
 At the high halo mass limit when $L_{\rm tot}(M,z) \gg L_{\rm cen}(M,z)$, halo luminosity dominated by satellites,
with $\gamma(M)$ a constant, we expect $N_{\rm sat}(M) \propto L_{\rm tot}(M,z) \sim M^{\beta_s+\alpha}$,
where $\beta_s$ is the slope introduced in equation~3 and $\alpha \sim 0.2$ 
is the slope of the $L_c(M)$ relation  at large halo masses.

\begin{figure*}[t]
\centerline{\psfig{file=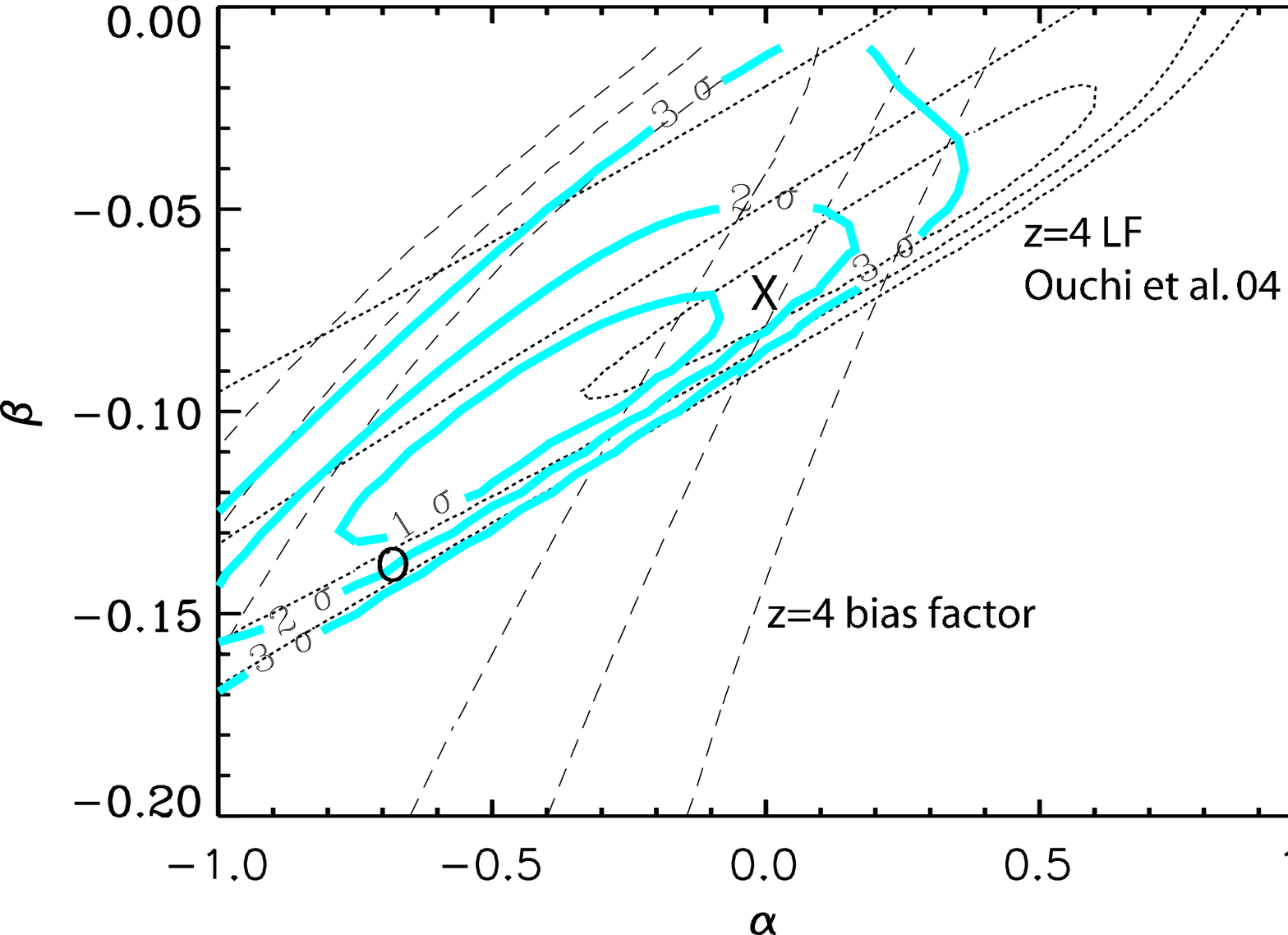,width=\hssize,angle=0}
\psfig{file=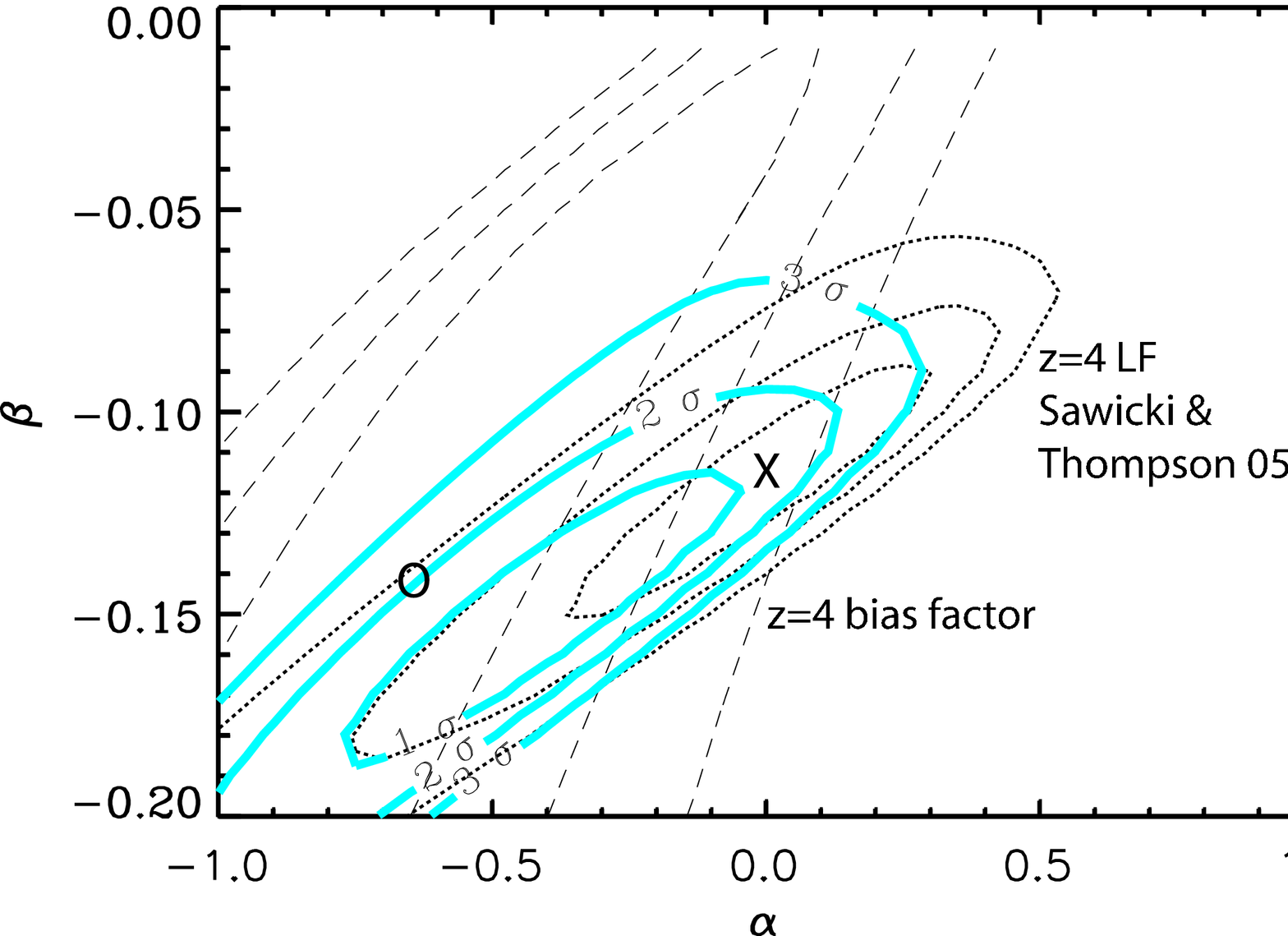,width=\hssize,angle=0}}
\caption{
Constraints on $L_c(M,z)$ evolution parameters, $\alpha$ and $\beta$ (see, equation~2),
where we plot $1\sigma$, $2\sigma$, and 3$\sigma$ allowed regions.
In both panels, the dashed lines show constraint from the luminosity-dependent  bias factor (from Figure~2), while
dotted lines show constraints from two estimates of $z=4$ LBG LF; The left panel uses Ouchi et al. (2004a) and the 
right panel uses Sawicki \& Thompson (2005). 
The location marked with an ``X'' are best-fit values for $\alpha$ and $\beta$ using the LF alone, while
the locations marked with an ``O'' are best-fit values  of $\alpha$ and $\beta$
with the bias factor alone. The thick solid contours are the overall constraint when bias and LF are combined.
The difference in the faint-end slope of the two LFs
requires different best-fit values to $\beta$, with $\sim -0.1$  using the
Ouchi et al. (2004a) LF and $\sim$ -0.15 using the Sawicki \& Thompson (2005) LF.
The best-fit value for $\alpha$, at $\sim -0.5$, remains the same regardless of the assumed shape for the LF.}
\end{figure*}

While we show the occupation numbers as appropriate for $z \sim 4$ LBG sample, we suggest that
a more appropriate statistic to use for various comparisons is the probability distribution function for a LBG
with a luminosity $L$ to appear in a halo of mass $M$:
\begin{equation}
P(M|L,z)dM=\frac{\Phi(L|M,z)}{\Phi(L,z)} \frac{dn(z)}{dM} \; dM \, ,
\end{equation}
where $\Phi(L,z)$ is the LBG LF given by
\begin{equation}
\Phi(L,z) = \int dM\, \frac{dn}{dM}(z)\, \Phi(L|M,z) \, ,
\end{equation}
and $dn/dM(z)$ denotes the mass function of dark matter halo.
Here we use the formalism of Sheth \& Tormen (1999) in our numerical calculations.
This mass function is in better agreement with numerical simulations (Jenkins et al. 2001)
when compared to the Press-Schechter mass function (Press \& Schechter 1974). This probability can be
divided to consider central and satellite galaxies separately.

Since clustering measurements are in terms of the angular correlation function,
we average over the galaxy redshift distribution associated with clustering measurements such that
\begin{equation}
w_p(\theta|L,z) =\int dr n^2(r) \int \frac{k dk}{2\pi} P(k|L,z) J_0(k d_A \theta) \, ,
\end{equation}
where $n(r)$ is the normalized radial distribution of galaxies with $\int dr n(r)=1$. 
Here $d_A$ is the comoving angular diameter distance. 

In the above, the power spectrum of galaxies $P(k|L,z)$ at a given luminosity
can be written in terms of the CLF
in terms of the 1- and 2-halo terms (see, review in Cooray \& Sheth 2002) at a redshift $z$ as
\begin{eqnarray}
&& P_\gal(k|L,z) = P_{1h}(k|L,z) + P_{2h}(k|L,z)\, ,
               \qquad{\rm where}\nonumber\\
&& P_{1h}(k|L,z) = \frac{1}{\bar{n}^2(L,z)}\int dM \, \frac{dn(z)}{dM} \\
&\times& \Big[\Phi^2_{\rm sat}(L|M,z)(L|M,z) u^2_{\rm gal}(k|M,z) \nonumber \\
&+& 2\Phi_{\rm cen}(L|M,z)\Phi_{\rm sat}(L|M,z)  u_{\rm gal}(k|M,z)  \Big]\nonumber \\
&& \quad \quad {\rm and} \nonumber \\
&& P_{2h}(k|L,z) = P^\lin(k,z) \Big[I_{\rm cen}(k|L,z) + I_{\rm sat}(k|L,z) \Big]^2 \, ,\nonumber \\
 \label{eqn:gal}
\end{eqnarray}
with the integrals $I_{\rm cen}(k|L,z)$ and $I_{\rm sat}(k|L,z)$ given  by
\begin{eqnarray}
&&I_{\rm cen}(k|L,z) = \int dM\, \frac{dn(z)}{dM} b_1(M,z) \frac{\Phi_{\rm cen}(L|M,z)}{\bar{n}(L,z)} \quad {\rm and} \nonumber \\
&&I_{\rm sat}(k|L,z) = \int dM\, \frac{dn(z)}{dM} b_1(M,z) \frac{\Phi_{\rm sat}(L|M,z)}{\bar{n}(L,z)} u_{\rm gal}(k|M,z) \, , \nonumber \\
\end{eqnarray}
respectively.
Here, and above,
\begin{equation}
 \bar{n}(L,z) = \int dM \, \frac{dn(z)}{dM}\, \left[\Phi_{\rm cen}(L|M,z)+\Phi_{\rm sat}(L|M,z)\right]
 \label{eqn:barngal}
\end{equation}
denotes the mean number density of LBGs while
\begin{equation}
 u_{\rm gal}(k|M,z) = \int_0^{r_{vir}} dr\ 4\pi r^2\,{\sin kr\over kr}\
{\rho_{\rm gal}(r|M,z)\over M} \, ,
\label{eqn:yint}
\end{equation}
denotes the normalized Fourier transform of the galaxy density distribution within
a halo of mass $M$ when $b_1(M,z)$ is the first-order bias factor of dark matter halos.

Here for dark matter halo bias we use the bias factor derived by
Sheth, Mo \& Tormen (2001) which corrects earlier
calculations by Mo et al. (1997; Efstathiou et al. 1988; Cole \& Kaiser 1989)
based on spherical collapse arguments. Here, we assume galaxies trace dark matter and
calculate the profile based on the NFW dark matter density profile (Navarro et al. 1997).
The concentration parameter is defined following the scaling relation of Bullock et al. (2001).
The relevant expressions in our calculation are summarized in Cooray \& Sheth (2002).
While the above expressions are written as a function of luminosity, since measurements
are made over a luminosity bin, we also calculate the predictions over a
luminosity bin rather than at a single luminosity. This is done by integrating over the
luminosity bin at which the measurements are made, assuming galaxies are uniformly
distributed over that luminosity bin.

\begin{figure}
\centerline{\psfig{file=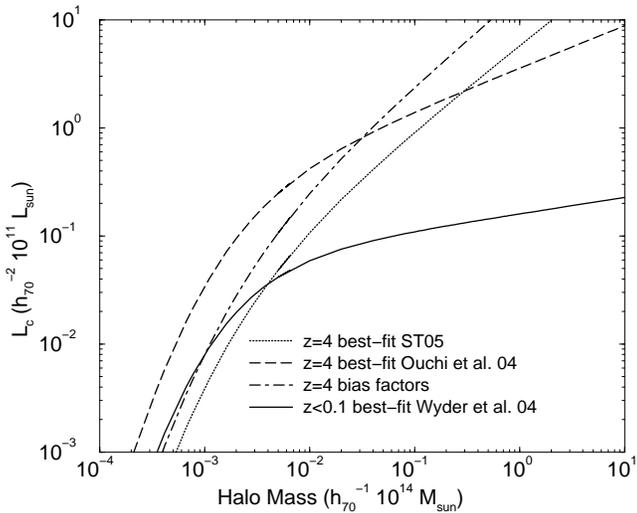,width=\hssize,angle=-90}}
\caption{
Central galaxy luminosity as a function of the
halo mass as appropriate for Far-UV band at $z < 0.1$ (solid line; as required to
describe the GALEX LF from Wyder et al. 2005), and at
$z \sim 4$ (in rest UV-bands). We show several examples based on best-fit values for
$\alpha$ and $\beta$ from Figure~4 (locations marked by ``X''s and  ``O'').
An increase in the rest-frame UV luminosity at
$z\sim4$ relative to today is clear. This redshift dependence can be described as a steepening of the
slope of the $L_c(M,z)$ relation at masses above 10$^{12}$ M$_{\sun}$,
while the faint-end slope remains the same. }
\end{figure}

At large physical scales, the galaxy power spectrum or the cross-power spectrum,
 reduces to that of the linear power spectrum scaled by a constant
galaxy bias factor. One can understand this by noting that
at large scales, $u_\gal(k|M,z)\to 1$ and the galaxy power spectrum simplifies to
\begin{equation}
 P_{\gal}(k|L,z) \approx b^2(L,z)\, P^\lin(k,z),
\end{equation}
where
\begin{eqnarray}
&& b(L,z) = \\
&&\int dM\, \frac{dn(z)}{dM}\, b_1(M,z)\,
          \frac{\left[\Phi_{\rm cen}(L|M,z)+\Phi_{\rm sat}(L|M,z)\right]}{\bar{n}_i(L,z)} \, , \nonumber
\end{eqnarray}
is  the mean large-scale bias factor.
This large-scale bias factor has already been discussed  with CLFs (see, Cooray \& Milosavljevi\'c 2005b; Cooray 2005b),
and we measure this bias as a function of LBG luminosity at $z \sim 4$ here.

\begin{figure*}
  \centerline{\psfig{file=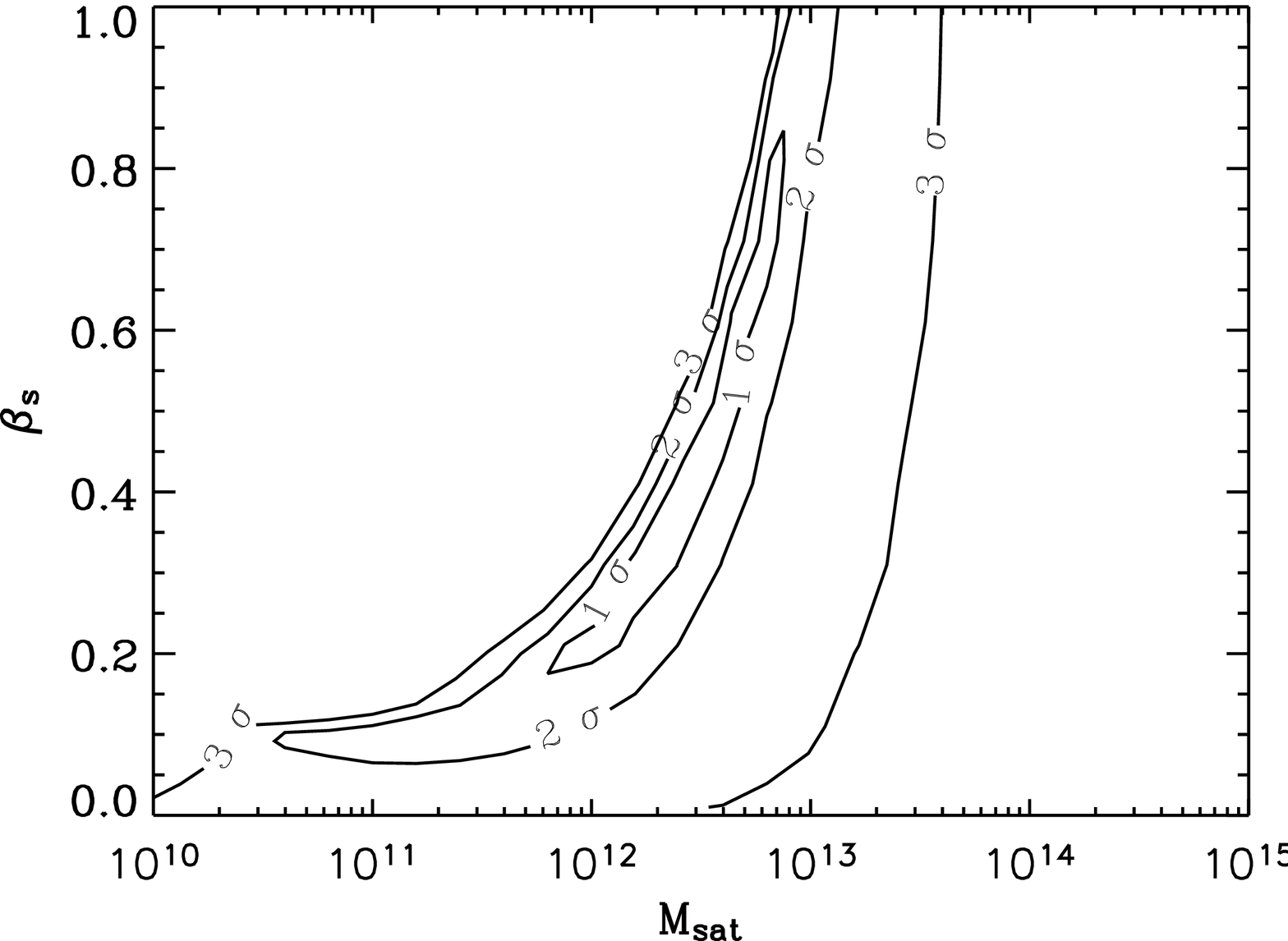,width=\hssize,angle=0}
\psfig{file=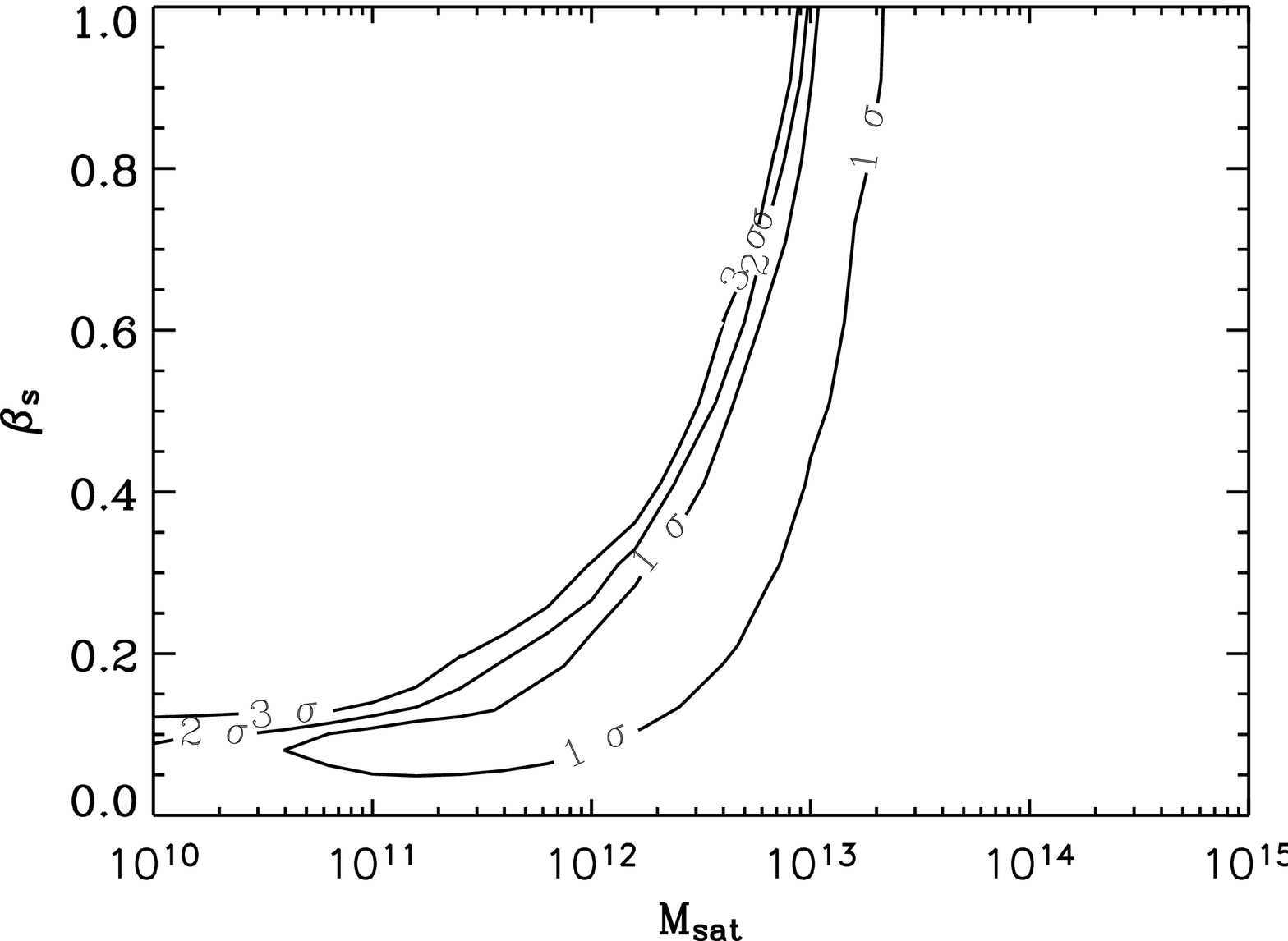,width=\hssize,angle=0}}
\centerline{\psfig{file=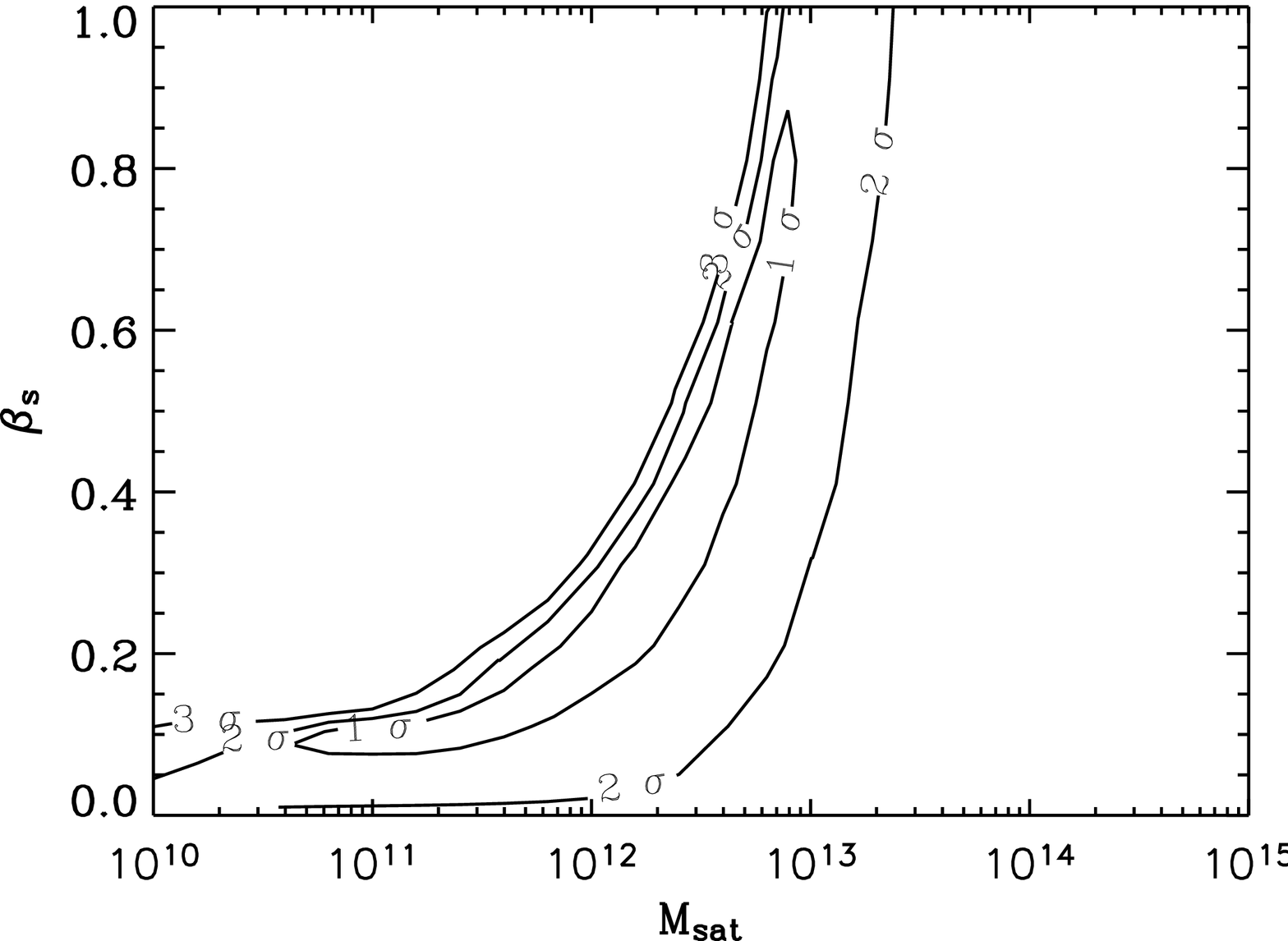,width=\hssize,angle=0}
\psfig{file=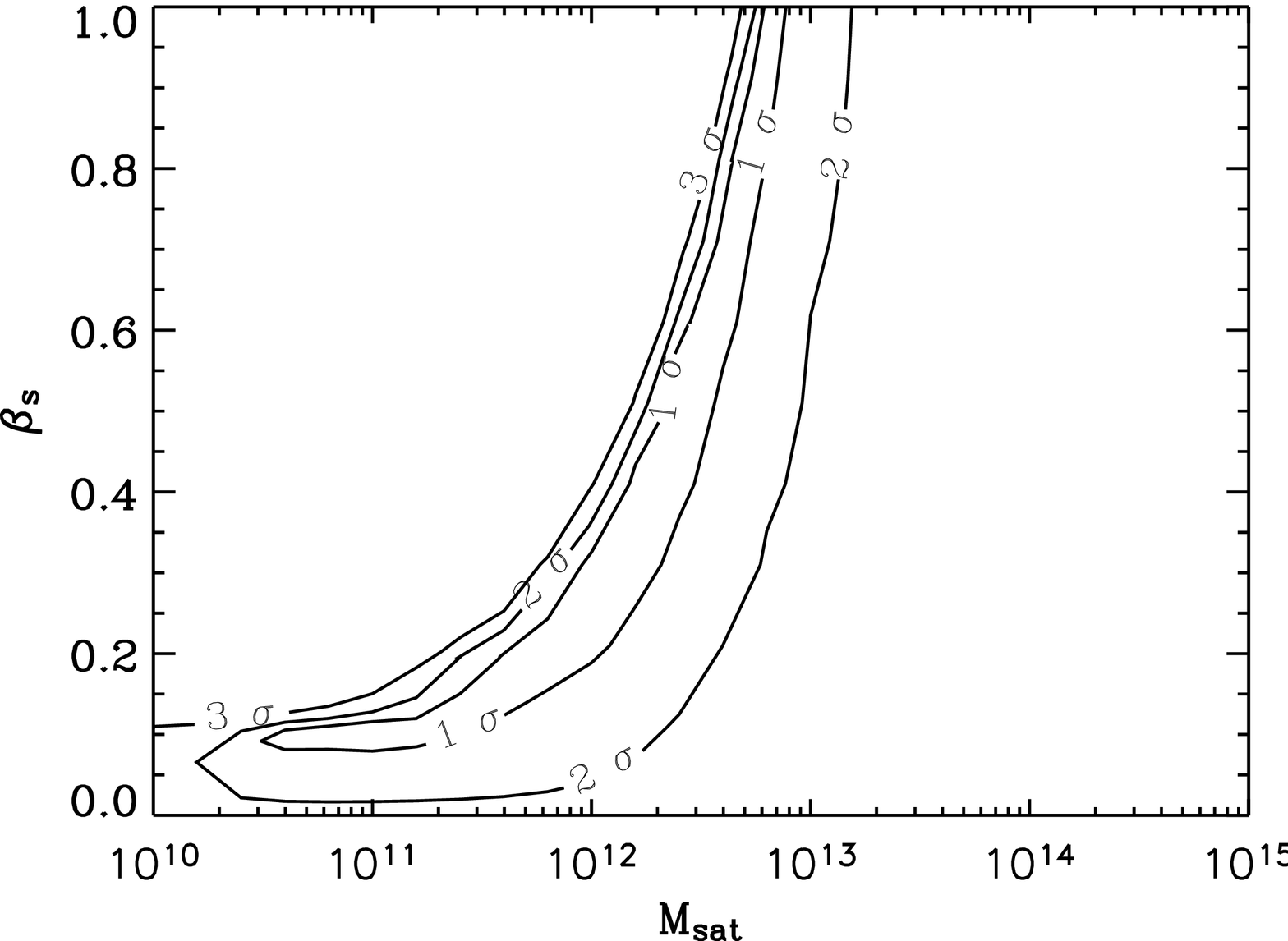,width=\hssize,angle=0}}
\caption{Constraints on parameters $\beta_{\rm s}$, the power-law slope of total luminosity--halo mass relation 
and $M_{\rm sat}$ (in units of $h_{70}^{-1}$ M$_{\sun}$), the halo mass scale at which satellites begin to appear,
related to the satellite CLF. The shown constraints are for
whole galaxy sample (top left), for galaxies with $24.5 < i < 25.5$ (top right),
 $25.5 < i < 26.5$ (bottom left), and $26.5 < i < 27.5$ (bottom right), respectively.}
\end{figure*}

\section{Results}

In Fig.~1, we show LBG autocorrelation function for the whole galaxy sample, with $i' < 27.5$,
and for subsamples divided to galaxy magnitude bins. The total autocorrelation function was measured and described in
Ouchi et al. (2005), while the luminosity dependent autocorrelation functions were measured as part of this study following
the description given in Section~2. In Fig.~1, for comparison, we also show model predictions which make use
of the best-fit parameters following a likelihood analysis we performed on the data. Here, $\beta_s$ and $M_{\rm sat}$ is related to
equation~2 and describes the satellite contribution to the CLF in the form of the total galaxy luminosity. 

The other ingredients used in our model fits are the $z=4$ LBG LF (Figure~2) and the luminosity-dependent bias factor (Figure~3).
We calculate  the bias factor at large scales by taking the ratio  of measurements 
to the linear theory prediction with a bias factor of unity, and establishing the mean of the ratio and the variance of this mean.
For this purpose, we only consider data  at angular scales greater than 10$''$ in Fig.~1 right panel. These bias factors
do not change significantly when we include the largest angular separation $w(\theta)$ measurement.
For reference, we also plot $z<0.1$ bias factor, as a function   of $M_r$ luminosity based on SDSS as measured  by Zehavi et al. (2005),
and the best-fit model description for low-redshift luminosity-dependent bias factors in SDSS from Cooray (2005c).  

For the likelihood analysis, we follow the  procedure described in Cooray (2005b) where the high-redshift rest B-band
LFs were analyzed in the context of CLFs. Here, we reproduce the same approach for rest UV LFs.
The $z=0$ $L_c(M)$ relation is  the one that reproduces the Far-UV LF from GALEX data 
(from Wyder et al. 2005). To establish this relation, we have followed the same procedure as Vale \& Ostriker (2004).  
The $z=4$ LF and LBG bias factor data are jointly fitted by varying 
parameters related to the $L_c(M,z)$  relation while keeping the same model description as the $z=0$ $L_c(M)$ relation, but modified
for redshift variation (equation~2).
Here, we consider a two-parameter model fit  by varying $\alpha$ and $\beta$. 
In these parameters, $\alpha$ varies the overall normalization while $\beta$ allows a mass-dependent variation to the $L_c(M,z)$
relation relative to the function today, $L_c(M,z=0)$. If $\alpha$ is non-zero while $\beta=0$, the $L_c(M,z)$ relation
and the LF behave such that one obtains a mass-independent pure luminosity evolution.
While $\alpha=0$ and $\beta=0$ imply the high redshift relation is same as the one today,
a non-zero value for $\beta$ would indicate that galaxy luminosities evolve in a way that the evolution depends on
the halo mass in which these galaxies appear.   

The parameter constraints
are shown in Fig.~4  with the left panel using the Ouchi et al. (2004a) LF and the right panel using the Sawicki \& Thompson (2005) LF. 
While both Ouchi et al. (2004a) and Sawicki \& Thompson (2005) LFs
are  consistent with $\alpha=0$, a value for $\beta$ of 0 is ruled at more than 3$\sigma$ confidence.
It is likely that the rest-UV LF
evolves in a complicated manner and a simple description based on pure luminosity evolution, where one shifts the
low redshift LF to high redshifts so as to make all galaxies brighter by the same factor,
is not appropriate.   An analysis of the rest B-band LFs from $z=0$ to $z\sim3$ revealed
a similar result in Cooray (2005b). 

In Fig.~5, we show $L_c(M,z)$ relation which makes the
mass-dependent evolution clear;
Central galaxies in $10^{12}$ M$_{\sun}$ or above dark matter halos are 
brighter at $z=4$ by a factor between 3 to 8  when compared to the luminosity of galaxies in same mass halos today.
Such a mass dependent increase in luminosity
at high redshifts is consistent with the ``down-sizing'' picture that has been regularly discussed in the context
of galaxy evolution (Cowie et al. 1996).

While LFs and galaxy bias allow model fits to the central galaxy CLF,
we now consider clustering of LBGs at $z \sim 4$ and fit the autocorrelation function
where non-linear clustering is determined by satellites.  We make use of the redshift distribution discussed in Section~2.
To study how our results are affected by uncertainties in the redshift distribution, we also
calculated parameter constraints by assuming that all galaxies are at a redshift of 4 and
determined that best-fit parameters change slightly, but is still below the 1$\sigma$ ranges
allowed by the data when including either the redshift distribution or assuming
all galaxies are at the same redshift. Thus, we do not consider uncertainties in the redshift
distribution to be a limitation of our analysis. When fitting the autocorrelation function,
we vary two parameters that describe the
appearance of satellite galaxies ($M_{\rm sat}$ and $\beta_s$) in Equation~3. 
We only consider measurements with $\theta < 300''$ for reasons described earlier in the text.

\begin{figure}
\centerline{\psfig{file=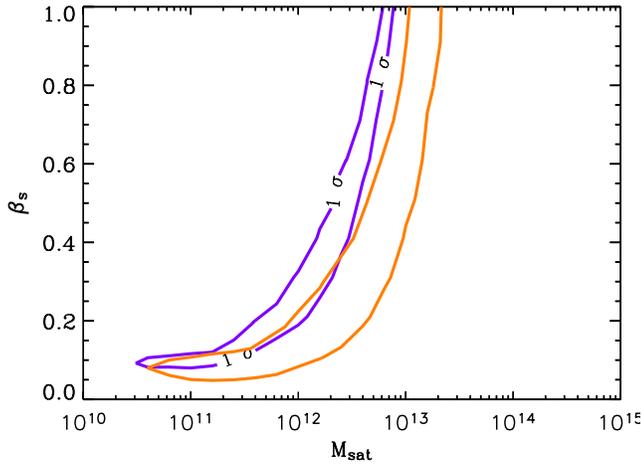,width=\hssize,angle=0}}
\caption{The difference between $24.5 < i < 25.5$ (bright end; contour to the left) and  $26.5 < i < 27.5$ 
(faint-end; contour to the right) LBG samples, where $M_{\rm sat}$ is in units of $h_{70}^{-1}$ M$_{\sun}$. 
For clarity, we only show the 1$\sigma$ constraint here, but there 
is evidence for the appearance of more luminous satellites in more massive halos. }
\end{figure} 

\begin{figure}
\centerline{
\psfig{file=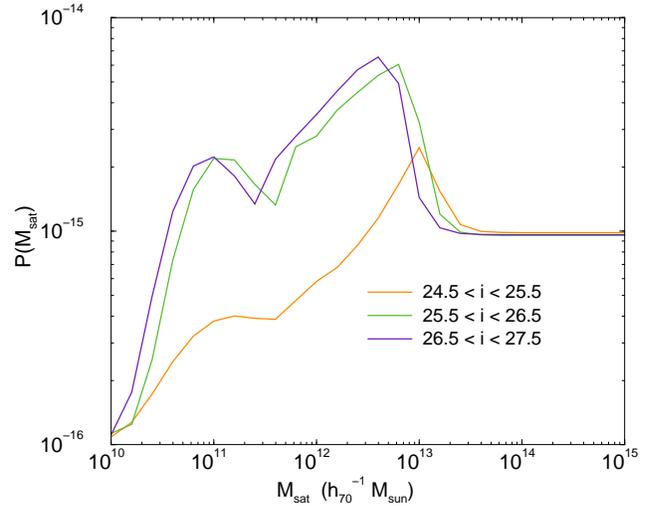,width=\hssize,angle=-90}}
\caption{Likelihood function for $M_{\rm sat}$ 
by marginalizing $\beta_s$ parameter, assuming uniform values between 0 and 1, as a function of the
magnitude bin.}
\end{figure}

Our results are summarized in Fig.~6 where we show constraints for the whole sample
(top left) and for autocorrelation functions divided to LBG luminosity bins.
Marginalizing over $\beta_s$, assuming uniform values between 0 and 1, we find
$M_{\rm sat}=3.9^{+4.1}_{-3.5} \times 10^{12}$ $M_{\sun}$, 
$6.2^{+3.8}_{-4.9}\times 10^{12}$ $M_{\sun}$, and $9.6^{+7.0}_{-4.6} \times 10^{12}$ $M_{\sun}$ (in units of $h_{70}^{-1}$ M$_{\sun}$).
for $26.5 < i' < 27.5$, $25.5 < i' < 26.5$,  and
$24.5 < i' <25.5$ magnitude bins, respectively, where error bars are 1$\sigma$ errors based on the likelihood analysis.

In Fig.~7, we summarize the one-sigma error contours, while in Fig.~8 we show the
 probability distribution functions based on the likelihood
analysis for $M_{\rm sat}$ for three LBG magnitude bins  considered in this study
by marginalizing over $\beta_s$ assuming a uniform prior between 0 and 1.
In Figs.~7 and 8 results are shown using the $L_c(M,z)$ relation based on the  Ouchi
et al. (2004a) LF; When constraints from the Sawicki \& Thompson (2005) LF is used, we found best-fit values of
$\beta_s$ and $M_{\rm sat}$ to vary by, at most, 20\% especially in the faintest luminosity bin.
While there is a general trend for $M_{\rm sat}$ to increase with increasing luminosity of 
galaxies, this trend is not well established given the overlapping error bars of $M_{\rm sat}$ estimates.  If $\beta_s > 0.4$, the trend becomes clear
with a difference in $M_{\rm sat}$ amounting to a factor of 4 to 5 between the faintest and the brightest luminosity bin.
However, there is no reason to restrict $\beta_s$ to high values, though additional clustering measurements
will allow parameters related to the satellite CLF be improved.

On the other hand, a clear trend exists for the fraction of galaxies
that appear as satellites in the three luminosity bins considered here. 
We calculate the satellite fraction via
\begin{equation}
f_{\rm sat}(L) = \frac{\int dM \Phi^{\rm sat}(L|M,z) \frac{dn(z)}{dM}}{\Phi^{\rm cen}(L,z)+\Phi^{\rm sat}(L,z)} \, .
\end{equation}
In Figure~9, as an example, we show contours of constant $f_{\rm sat}$ values as a function of $M_{\rm sat}$ and $\beta_s$
overlaid on the constraints from model fits to galaxy clustering in $26.5 < i < 2.75$ magnitude bin.
The satellite fraction,  $f_{\rm sat}$, traces
the degeneracy direction in the $M_{\rm sat}$---$\beta_s$ plane and is a parameter that is slightly better determined from the data.

In Figure~10, we show probability distribution for $f_{\rm sat}$ based again on the likelihood analysis.
The fraction of galaxies that appear as satellites is $0.26^{+0.04}_{-0.07}$, $0.18^{+0.04}_{-0.05}$,
and $0.09^{+0.05}_{-0.06}$ for $26.5 < i' < 27.5$, $25.5 < i' < 26.5$,  and $24.5 < i' <25.5$  magnitude bins
 respectively, if one uses constraints on the $L_c(M,z)$ relation from Ouchi et al. (2004a) LF.
If Sawicki \& Thompson (2005) LF is used, the satellite fractions are lower with values
$0.20^{+0.06}_{-0.07}$, $0.14^{+0.05}_{-0.05}$,
and $0.08^{+0.06}_{-0.05}$ for $26.5 < i' < 27.5$, $25.5 < i' < 26.5$,  and $24.5 < i' <25.5$  magnitude bins, respectively.
The difference, however, exists mostly in the lowest luminosity bin with a 25\% change in the best-fit value, 
while at the highest luminosity bin,
fractions from the two LFs agree with each other. This is probably a reflection that the two LFs agree at the bright-end
and the differences in the two LFs only exist at the faint-end.
In both cases, there is a clear 
trend in $f_{\rm sat}$ with luminosity such that the fraction  of satellites  is smaller for the brighter LBG sample than
that for the fainter sample.

\begin{figure}
\centerline{\psfig{file=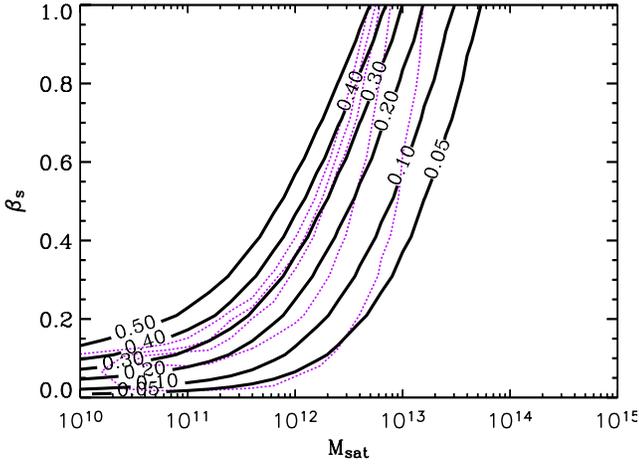,width=\hssize,angle=0}}
\caption{
The fraction of galaxies with  $26.5 < i < 27.5$ magnitudes
that appear as satellites, $f_{\rm sat}$
as a function of $\beta_{\rm s}$, the power-law slope of total luminosity--halo mass relation,
and $M_{\rm sat}$ (in units of $h_{70}^{-1}$ M$_{\sun}$), the halo mass scale at which satellites appears.
For reference, in thin dotted lines, we overlap the constraints on this parameter space from LBG clustering data (from Fig.~6).
}
\end{figure} 

\begin{figure}
\centerline{\psfig{file=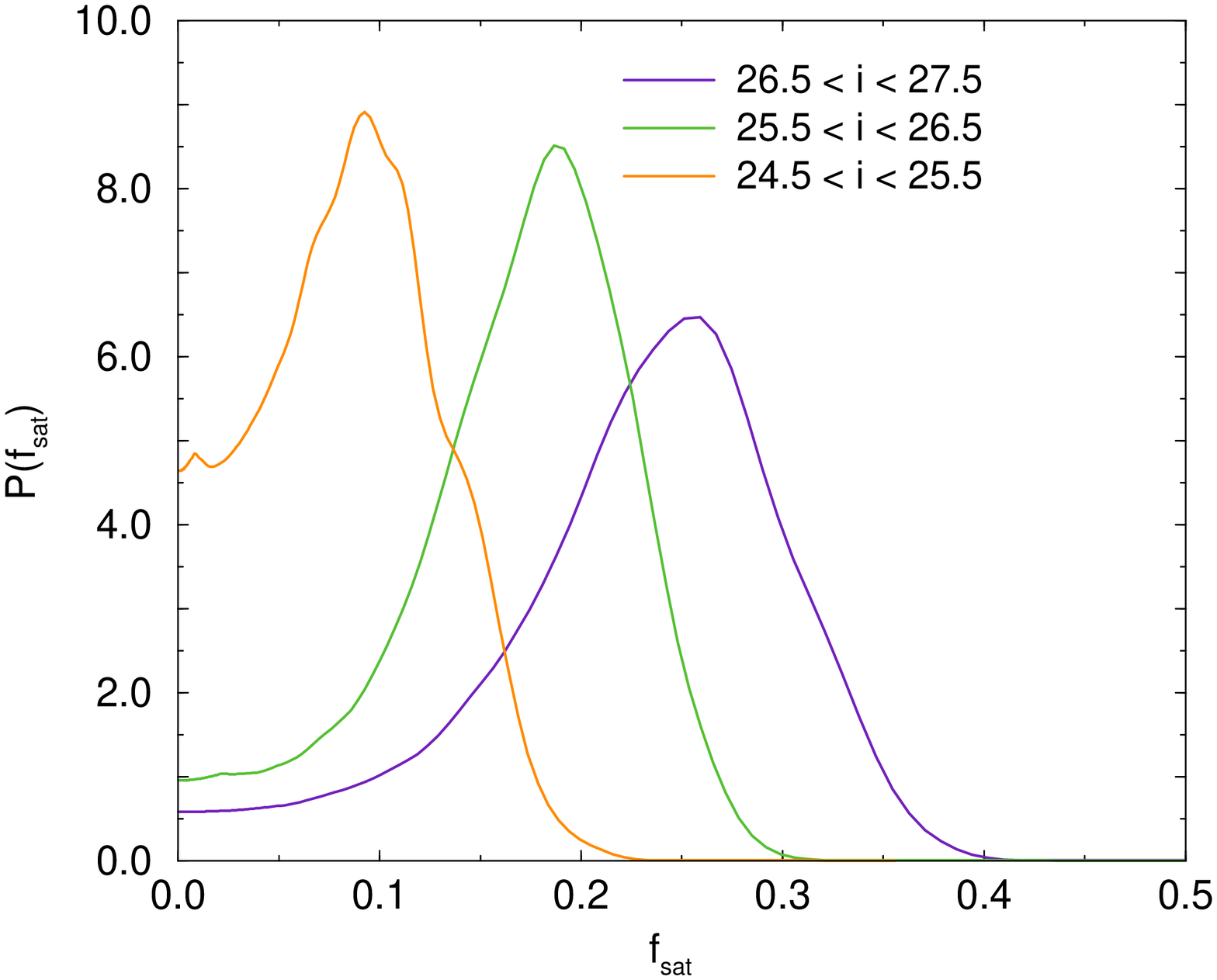,width=\hssize,angle=-90}}
\centerline{\psfig{file=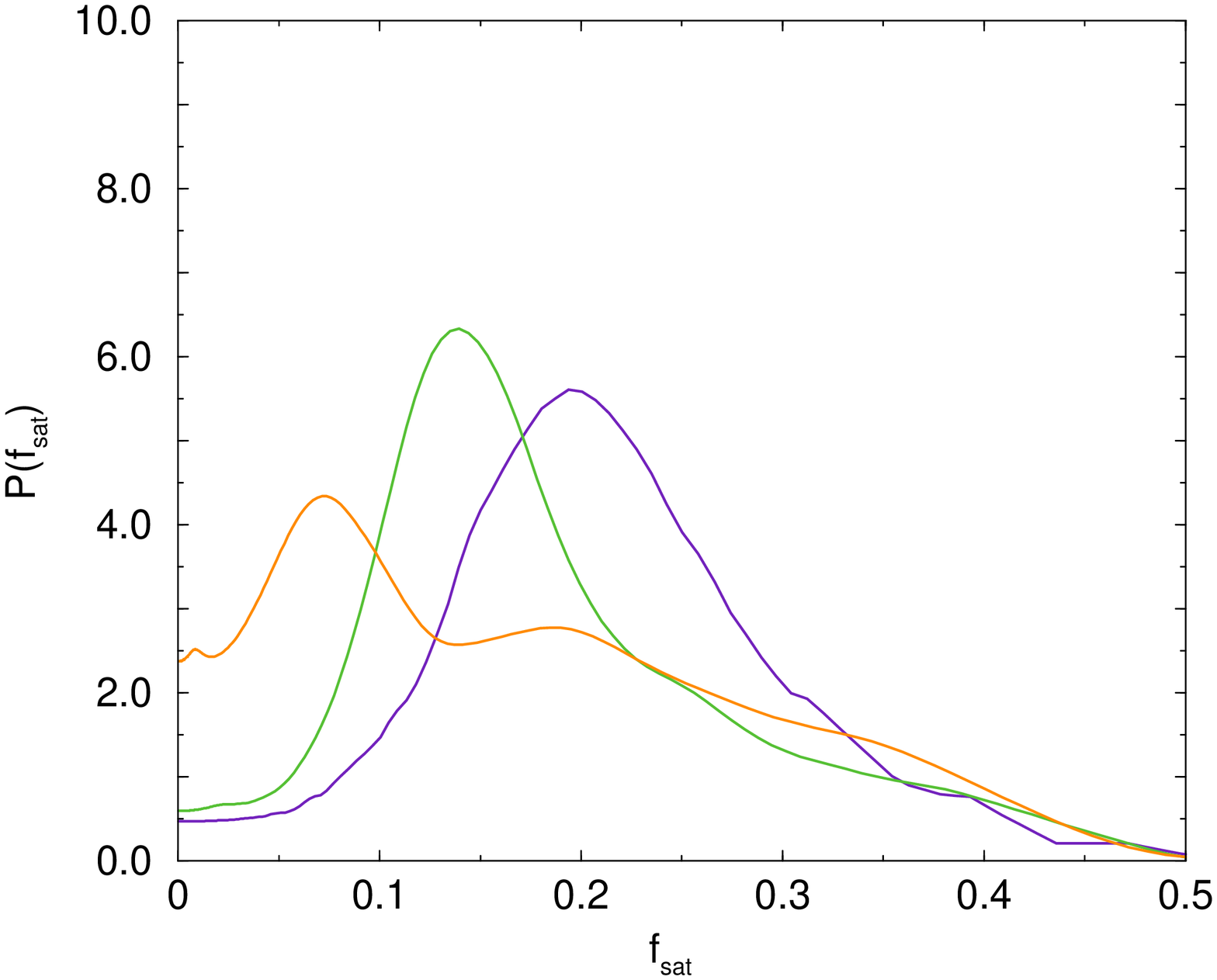,width=\hssize,angle=-90}}
\caption{The probability distribution of $f_{\rm sat}$, the satellite fraction at a given luminosity bin, based on model fits to the data.
The top panel shows the fraction based on the Ouchi et al. (2004a) LF, while the bottom panel uses the Sawicki \& Thompson (2005) LF.
The latter LF has a flat slope at the faint-end leading to a slightly lower fraction of galaxies that appears
as satellites than the result based on the Ouchi et al. (2005) LF involving a steep slope at the faint-end.}
\end{figure} 

\begin{figure*}
\centerline{
\psfig{file=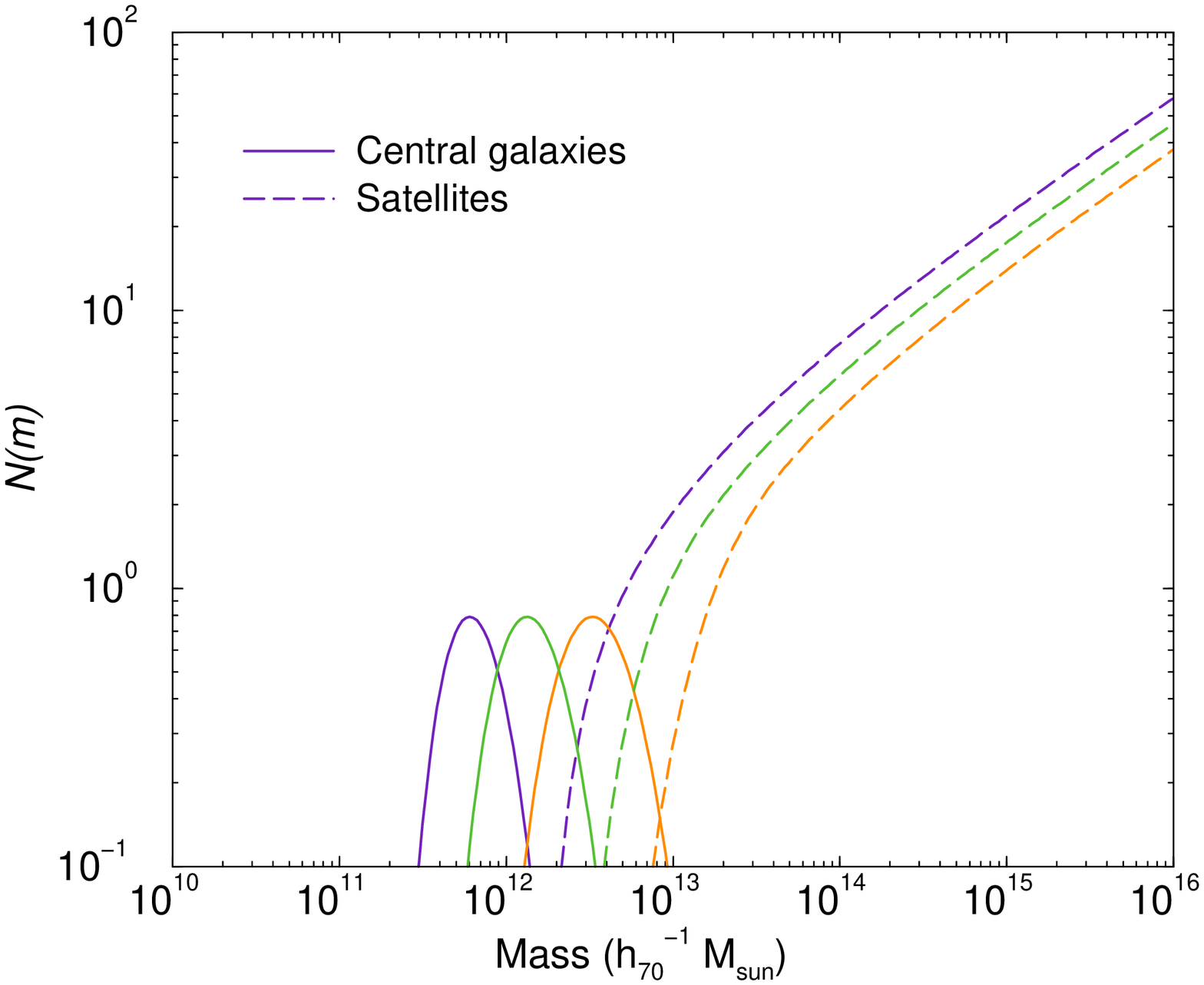,width=\hssize,angle=-90}
\psfig{file=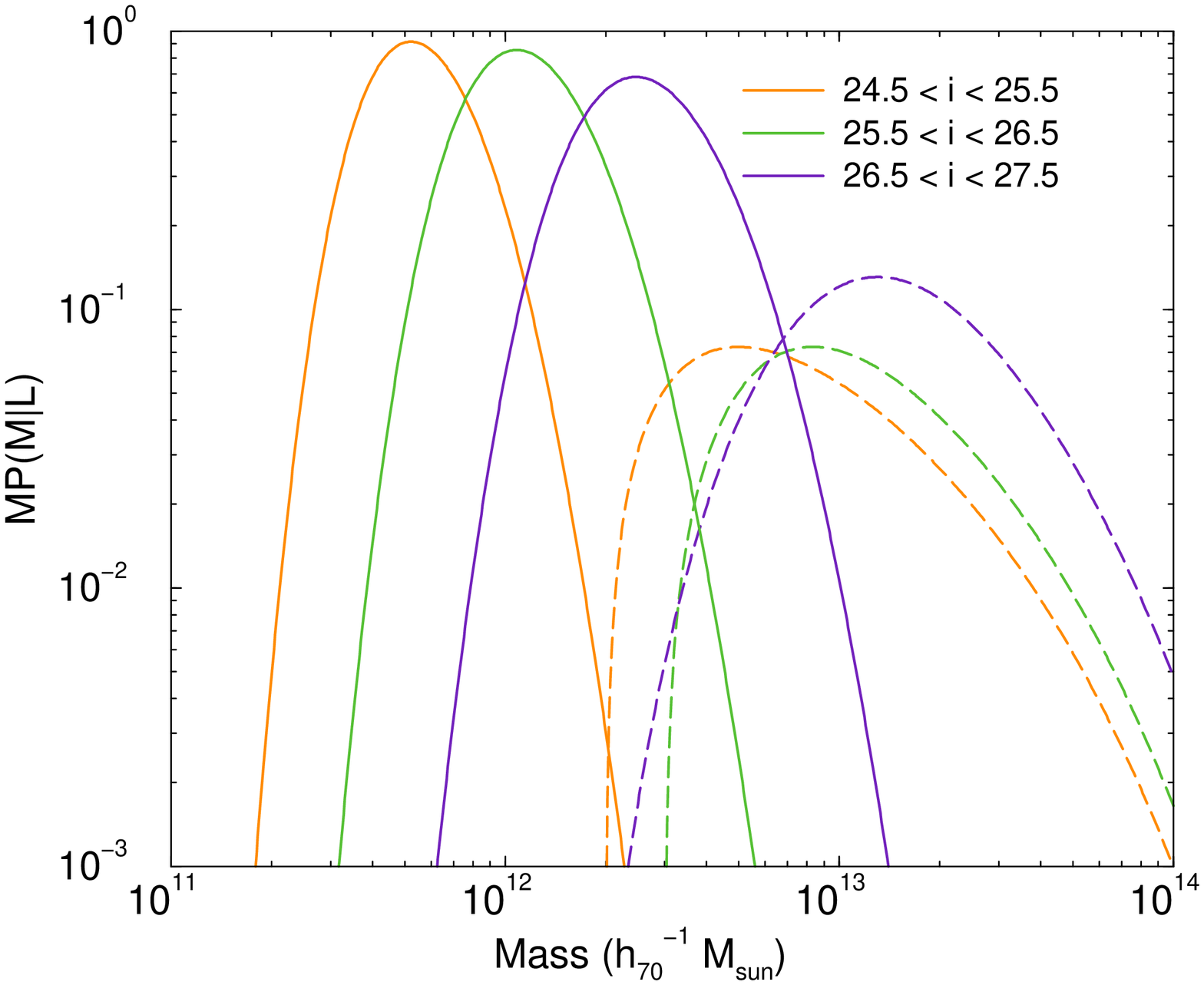,width=\hssize,angle=-90}}
\caption{{\it Left panel:} The halo occupation numbers as a function of the galaxy luminosity. Here,
we only show the best-fit parameter models for the satellite CLF here, though there are large variations
in both the slope and the mass scale in which satellites begin to appear.
The dashed-lines here show the probability distribution for satellite galaxies while solid lines are for
central galaxies as a function of the luminosity bin indicated on the panel.
{\it Right panel:} The probability distribution function of halo mass to host galaxies in a given luminosity
bin. We show both central galaxy (solid lines) and satellite (dashed lines) probability
distribution functions. Note the clear dependence of luminosity with halo mass for central galaxies.}
\end{figure*}

In Fig.~11 left panel, for reference, we show the best-fit halo occupation numbers as a function of the galaxy luminosity. Here,
we only use the best-fit parameters when describing the satellite CLF here, though there are large variations
in both the slope of the luminosity distribution of satellite galaxies and the mass scale in which satellites first begin to appear
at a given luminosity.
In the right panel of Fig.~11, we also show the probability distribution function of halo mass to host galaxies in a given luminosity
bin (following equation~5). We show both central galaxy (solid lines) and satellite (dashed lines) probability
distribution functions. Note the clear dependence of luminosity with halo mass for central galaxies.
These probability distribution functions can be compared with the ones derived for low-redshift galaxies in Cooray (2005c).
In fact, a trend can be established for central galaxies at a given luminosity to evolve
in halos of varying dark matter mass as a function of redshift. This 
trend is such that the halo mass increased as the redshift is lowered
for galaxies with high luminosities.
The difference between the halo masses of faint central galaxies as a  function of redshift
is not significant. These differences can be understood in terms of the difference in $z < 0.1$ and
$z \sim4$ $L_c(M,z)$ relations (shown in Fig.~5). It could be that these 
trends can be explained through hierarchical merger models that trace the
merger rates of dark matter halos from high redshift to low redshifts (e.g., Hamana et al. 2005) and such
models can be further improved with more accurate clustering measurements as a function of redshift and luminosity.

In addition to simply establishing the mass scale, the probability distribution functions shown in Fig.~11 can be
compared with dynamical mass estimates for same galaxies to study the importance of merger-bias
(Furlanetto \& Kamionkowski 2005) and corrections associated with age dependences to the bias (e.g., Gao et al. 2005).
The lack of dynamical mass estimates for $z=4$ LBG sample, however, prohibits us from making such a comparison.
In addition to dynamical measurements, another useful avenue to explore would be 
a lensing-based mass measurement for LBG galaxies. While
an adequate surface density of lensed background galaxies behind LBGs are unlikely to be recovered even in
deep imaging data, a combination of the lensed cosmic microwave background anisotropy map at high resolution
and foreground LBG samples binned in luminosity may make such an analysis eventually possible.

\section{Summary and Conclusions}

To summarize our discussion involving model descriptions of luminosity-dependent clustering at $z\sim 4$, our main results are:

(1)  We have remeasured galaxy autocorrelation function in the $z\sim 4$ LBG sample of the Subaru/XMM-Newton Deep Field. While
Ouchi et al. (2005) considered galaxy clustering as a function of the increasing faint-end magnitude, while the bright-end
is fixed the same, here we have considered three independent bins in luminosity. We have made high signal-to-noise ratio
measurements of clustering in three luminosity bins from 24.5 to 27.5 in observed $i'$-band at steps of unity in  magnitude.

(2) In addition to these clustering measurements and the linear large-scale bias factor for LBGs
as a function of the luminosity derived from the data, we also analyze the $z\sim 4$ LBG LF from Ouchi et al. (2004a)
and Sawicki \& Thompson (2005). We make use of these two luminosity functions as they span
a wide range of possibilities in the faint end from a steep slope, in the case of Ouchi et al. 2004a LF,
to a flat slope with Sawicki \& Thompson (2005) LF.
Our general model fits suggest a mass-dependent luminosity evolution scenario for central galaxies
such that galaxies that are present in halos above $\sim 10^{12}$ M$_{\sun}$ brighten by factor of 3 to 8 between now and $z \sim 4$. 
This suggests that the star formation rate,
per given dark matter halo mass,  was increasing to high redshifts from today.
A conclusion similar to what we generally suggest here was also
reached by Dahlen et al. (2005) based on the rest B-band LF constructed from GOODS data, and also by Cooray (2005b)
based on LFs from DEEP2 (Willmer et al. 2005), COMBO-17 (Wolf et al. 2003), and from compilations of
Gabasch et al. (2004) and Giallongo et al. (2005) as a function of redshift.

(3) We see a possible trend for more luminous galaxies to appear
as satellites in more massive halos: Assuming that slope of satellite luminosity--halo mass is
between 0 and 1, the minimum halo mass in which galaxies begin to appear as satellites is 
$3.9^{+4.1}_{-3.5} \times 10^{12}$ $M_{\sun}$,
$6.2^{+3.8}_{-4.9}\times 10^{12}$ $M_{\sun}$, and $9.6^{+7.0}_{-4.6} \times 10^{12}$ $M_{\sun}$
in $26.5 < i' < 27.5$, $25.5 < i' < 26.5$,  and
$24.5 < i' <25.5$ magnitude bins, respectively, where error bars are 1 $\sigma$ confidence errors based on a likelihood analysis.
Due to large uncertainties, however, the trend with luminosity is not well established from the data,
except to some extent in the two outer bins.

(4) A clear trend exists for the fraction of galaxies
that appear as satellites in the three luminosity bins considered here.  For example, if the steep faint-end LF  from Ouchi et al.
(2004a) is the correct description, the fraction of galaxies that appear as satellites is $0.26^{+0.04}_{-0.07}$, $0.18^{+0.04}_{-0.05}$,
and $0.09^{+0.05}_{-0.06}$ for $26.5 < i' < 27.5$, $25.5 < i' < 26.5$,  and $24.5 < i' <25.5$  magnitude bins
 respectively. With the Sawicki \& Thompson (2005) LF that shows a flat slope at the faint-end, 
these  fractions are
$0.20^{+0.06}_{-0.07}$, $0.14^{+0.05}_{-0.05}$,
and $0.08^{+0.06}_{-0.05}$ for $26.5 < i' < 27.5$, $25.5 < i' < 26.5$,  and $24.5 < i' <25.5$  magnitude bins, respectively.
There is clear dependence on the increasing fraction  of satellites with
decreasing LBG luminosity. This is probably the first instance that satellite fraction of $z \sim 4$ LBGs have been determined
with sufficient accuracy to see a trend with luminosity, though the fractions could be better established
with an improved determination of the $z =4$ LBG LF, when combined with luminosity-dependent clustering statistics.

{\it Acknowledgments:} 

We thank the Subaru/XMM Newton Deep Survey Team for providing the $z=4$ LBG catalog that made
this difficult analysis possible. We are grateful to Takashi Hamana for reading a draft of this
manuscript and for his comments.   AC thank Caltech for hospitality while this work was initiated
and acknowledge discussions with Milos Milosavljevi\'c and Andrey Kravtsov. 
MO is supported by NASA through Hubble Fellowship
grant HST-HF-01176.01-A awarded by the Space Telescope Science Institute,
which is operated by the Association of Universities for Research
in Astronomy, Inc., for NASA, under contract NAS 5-26555.

\end{document}